\documentclass[twocolumn,showpacs,prl]{revtex4}
\usepackage{graphicx}
\usepackage{natbib}

\newcommand{\ds}{\displaystyle}

\newcommand{\be}{\begin{equation}}
\newcommand{\ee}{\end{equation}}
\newcommand{\beq}{\begin{eqnarray}}
\newcommand{\eeq}{\end{eqnarray}}

\newcommand{\w}{\omega}
\newcommand{\W}{\Omega}
\newcommand{\g}{\gamma}
\newcommand{\G}{\Gamma}

\newcommand{\ket}{\rangle}
\newcommand{\bra}{\langle}

\newcommand{\bnn}{\begin{eqnarray*}}
\newcommand{\enn}{\end{eqnarray*}}

\newlength{\textwidthm}
\begin{document}

\title{
From EIT photon correlations to Raman anti-correlations in coherently 
prepared Rb vapor.
}
\author{
    Vladimir A. Sautenkov$^{1,2}$,
    Yuri V. Rostovtsev$^{1}$, and
    Marlan O. Scully$^{1,3}$
}

\affiliation {
    ${^1}$ %
Institute for Quantum Studies and Department of Physics, Texas A\,\&\,M University, 77843 \\
    ${^2}$Lebedev Institute of Physics, Moscow 119991, Russia\\
    ${^3}$ 
Princeton Institute for the Science and Technology of Materials
and Department of Mechanical \& Aerospace 
Engineering, Princeton University, 08544\\
%
%
}

\date{\today}

\begin{abstract}
\vskip12 pt
We have experimentally observed switching between 
photon-photon correlations (bunching) and anti-correlations (anti-bunching) 
between two orthogonally polarized laser beams in an EIT configuration 
in Rb vapor. The bunching and anti-bunching sswitching occurs at a specific 
magnetic field strength.

\end{abstract}

\maketitle


Interaction of macroscopic laser fields with a resonant atomic 
3-level Lambda system has been the focus of recent research. 
In particular, interesting phenomena such as  
photon-photon correlation, phase squeezing, and new entangled states of 
the radiation field have been studied~\cite{Kimble,Lukin,Harris05prl,Alzar2,Zubairy05prl}.
Sub-Poissonion statistics and a reduction of noise below photonic short 
noise have been demonstrated~\cite{Alzar2}. 

New bright source of entangled photon states with controlable 
coherent time has been demonstrated~\cite{Harris05prl} by employing 
parametric generation with 
counter-propagating electromagnetic waves in Rb vapor.
The photon statistics of the light emitted from atomic ensemble into a single
field mode of an optical cavity has been studied in \cite{Rempe05prl}, smooth
transition from bunching to anti-bunching has been experimentally 
demonstrated by changing number of atoms in the cavity. 

In this Letter, we study intensity correlations and anti-correlations 
of optical fields propagating through a dense Rubidium vapor. The main result
of the paper is shown in Fig.~1. For lower level coherence between 
Zeeman sub-levels prepared by laser beams with orthogonal polarizations 
(Fig.~1a),
we observe that the intensity fluctuations of two laser beams are correlated 
under the condition of electromagnetically induced transparency (EIT) 
(see in Fig.~1b) and anti-correlated when 
two-photon detuning is introduced (see in Fig.~1c). 
That is, we have observed an interesting transition between bunching and 
anti-bunching by changing two-photon detuning. The two-photon 
spectral width of this transition is narrower than the width of the EIT 
window. A theoretical approach similar to that developed in 
\cite{Scully,schleich} has been used to explain the observed results. 

\begin{figure}[tb] 
\center{
\includegraphics[width=8cm]{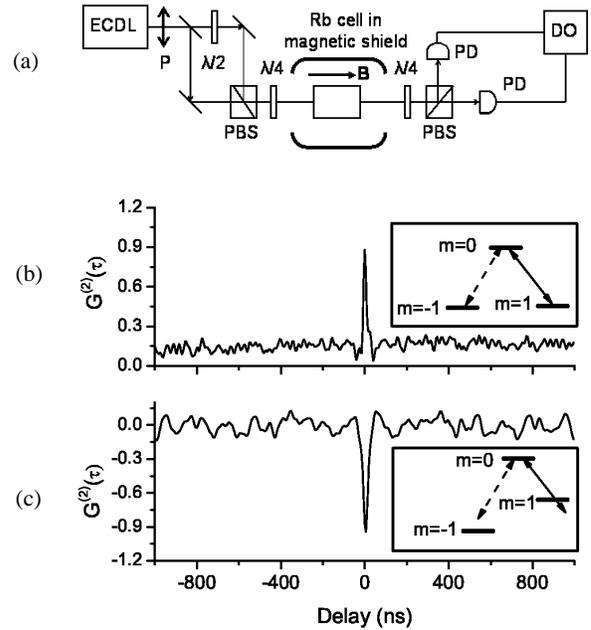} 
} 
\caption{\label{exp-scheme}
(a) Experimental setup. ECDL is the extended cavity diode laser, 
P is the polarizer, PBS is the polarizing beam splitter, PD is the photo diode,
DO is the digital oscillascope. 
Intensity correlation functions, $G^{(2)}(\tau)$, 
are shown corresponding to two cases: (b) no magnetic field, EIT, and (c) 
with magnetic field, $B=-0.47$ Gauss. It is shown in insets the simplified 
energy levels without and with magnetic field.
%
}
\end{figure}

%

\begin{figure} 
\center{
\includegraphics[width=7cm]{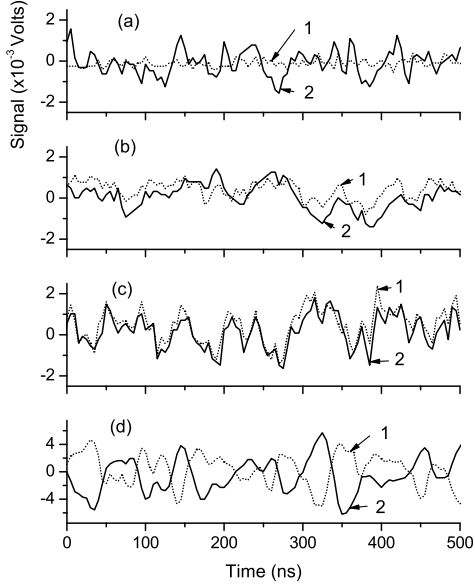}
} \caption{\label{Noise-signals} 
Waveforms from photo-detectors with optical power of
laser beams at front window of Rb cell is 0.5 mW.
The fluctuaions of intensity of 
(a) one beam (the second is blocked) before and after the cell, 
(b) two spatially separated beams, 
(c) two coinciding beams without magnetic field ($B=0$) 
at the EIT, zero two-photon detuning and 
(d) for magnetic field $B=-0.47$ Gauss. 
}
\end{figure}


\begin{figure} 
\center{
\includegraphics[width=7cm]{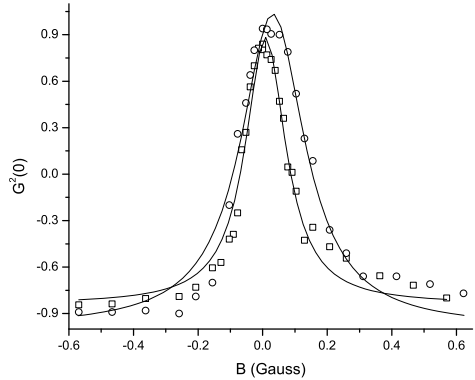}
} \caption{\label{Correlation vs B}
%
Correlation $G^{(2)}(0)$ versus magntic field (two-photon detuning). 
Circles  corresponds to optical power 0.5 mW, squares - optical power 0.25 mW.
}
\end{figure}


An experimental setup is shown in Fig.~1a. 
%
%
All measurements have been performed with an external-cavity diode
laser (ECDL) tuned to the center of the Doppler broadened $D_1$ line 
($5S_{1/2}$ $(F=2)$ $\leftrightarrow$ $5P_{1/2}$ $(F'=1)$). 
The output laser beam is split into two by a beam-splitter, 
then the polarization of the one of the beams
is rotated by a $\lambda/2$-wave plate, and these two orthogonally linear
polarized beams are combined together by a polarizing beam-splitter (PBS). 
After the $\lambda/4$ wave-plate the beam is a combination
of two orthogonal circularly polarized optical fields which induce
a ground state Zeeman coherence in Rb atoms. 
A simplified $\Lambda$-scheme is depicted in insets of Fig.~1(b,c).  
A glass cell ($l=7.5$ cm) with Rb vapor at density $10^{12}$ cm$^{-3}$
has been installed in a two-layer magnetic shield. 

After the cell, the transmitted laser beams with
opposite circular polarizations are separated by a second
$\lambda/4$ wave-plate and another polarizing beam splitter.
Transmitted laser beams were focused on fast photodiodes (PD) ET
2030A (Electro-Optics Technology) with frequency bandwidth $75$ kHz -
$1.2$ GHz. The optical path lengths for both beams are the same.
Signals from PDs are analyzed by a digital oscilloscope (DO) (Agilent
54624A with frequency bandwidth 100 MHz).
%



At the optical power of each beam at the entrance window of 
the Rb cell is 0.5 mW, beam diameters 0.1 cm, we vary a magnetude of 
longitudinal magnetic field and observe change in transmission. 
Linear transmission is less than 1\%, and at maximum of EIT transmission 
is 0.63. 
The measured FWHM is $0.86$ Gauss that is five times broader than 
for a low power limit $0.16$ Gauss, i.e.  
the width of EIT is determined by strong power broadening. 
By using ground state Zeeman splitting $\Delta\omega/2\pi_{B}=0.7$ MHz/Gauss,  
one can estimate the spectral width of the EIT resonance of 
1.2 MHz that are several times narrower than natural FWHM 
of hyperfine transitions 6 MHz and Rabi frequency.
%

To study fluctuations of optical fields transmitted through 
the dense Rb vapor, we have registered the dependence of 
the intensities of both optical beams on time, 
$\bra I_{1,2}\ket + \delta I_{1,2}(t)$. Here $\bra I_{1,2}\ket$ are the 
average intensities of the laser beams, and $\delta I_{1,2}(t)$ are the 
time dependent intensity fluctuations shown in Fig.2. 
Data presented in Fig. 2 are a part of the 10 $\mu$sec recorded data. 
The signal in Volts is proportional to laser intensity as 500 V/W. 
Waveforms shown in Fig.2a are obtained under the condition that when one of the
beams is blocked before the cell and there is no magnetic field $(B =0$). 
The first curve shows intensity fluctuations 
of optical beam before it goes into 
the cell, it has practically no amplitude noise, and in order to see 
optical intensity noise above the photodetector noise the optical power has 
to be increased at least three times. The second curve (with a larger  
waveform amplitude) shows excess noise of 
transmitted optical fields induced by the dense $Rb$ vapor. 

Waveforms in Fig.2b show intensity fluctuations of two separated optical
beams. Beams are separated by a tilted glass plate (an incident angle is 
close to the Brewster angle) with good quality parallel optical surfaces.
Distance between beams is 0.3 cm that is bigger than the beam diameters. 
One can see that the vapor induced noise is slightly correlated. 
Correlation increases when separation is reduced. Strong
correlation between signals is shown in Fig.2c under resonance
EIT conditions. In Figure~2d
waveforms are recorded with two-photon detuning by applying a magnetic 
field $B = -0.47$ Gauss.

One can see from Fig.2 that there are strong 
correlations and anticorrelations between beams governed by the 
EIT condition.   
The correlation function $G^{(2)}(\tau)$ between intensities of 
two optical beams is calculated by 
\be
G^{(2)}(\tau)=
\frac{\bra\delta I_{1}(t) \delta I_{2}(t+\tau)\ket}{
\sqrt{\bra[\delta I_{1}(t)]^2\ket\bra[\delta I_{2}(t+\tau)]^2\ket}
} 
\ee
where averaging over the time is defined as 
$\bra Q(t)\ket = \int_t^{t+T}Q(t)dt/T$, 
$\tau$ is the selected time delay between the recorded signals,
$T$ is the time of integration, in our case $T=10$ $\mu$s.

In Figure 1b, the curve is calculated by using
waveforms recorded under resonance EIT condition ($B =0$). The
correlation peak at delay time $\tau =0$ has an amplitude of $0.9$ and the
background average value near $0.15$. Figure 1c demonstrates
a pronounced modification of waveforms due to two-photon detuning
(applied magnetic field $B = -0.47$ Gauss) showing that the waveforms are
anti-correlated. For the reduced optical power, the width of 
the correlation peaks has decreased. The widths of the peaks 
is associated with the saturated width of resonance in Rb vapor 
absorption (a single photon resonance) \cite{saturation}.

To study the role of coherence and to ensure that the results obtained 
are related to the interaction between 
optical beams we perform experiments with separated beams. 
The results of analysis for power of $0.5$ mW show 
that the correlation $G^{(2)}(\tau)$ obtained with spatially separated beams
has the background 0.3 and a correlation peak with
magnitude 0.7 at the delay $\tau = 0$. 
When we reduce the spatial separation between beams, 
the contrast and the magnitude of the correlation pick is increased. 

We perfom a set of measurements of $G^{(2)}(0)$ for different
values of the magnetic field $B$. The correspondent dependence 
vs magnetic field for optical powers 0.5 mW and
0.25 mW are shown in Fig.3. At zero magnetic field ($B = 0$) the
correlation between $\delta I_{1}(t)$ and $\delta I_{2}(t)$ is close
to 0.9. The correlation decreases with increasing of the xmagnetic field, 
it goes through zero, and then it reaches magnitude $-0.9$. 

Intensity correlation $G^{(2)}(0)$ demonstrates a resonance-like 
behavior on the change of magnetic field. The widths of these resonances 
are $0.24$ and $0.16$ Gauss correspondingly that is nearly 4 times narrower 
than the width of the EIT window at the same optical power. 
The EIT width (FWHM) increases from $0.65$ Gauss to $0.86$ Gauss 
with optical power of the each laser beam 
from $0.25$ to $0.5$ mW. The smallest measured width of the correlation
resonance coincides with the width of the EIT resonance at the low power
limit $0.16$ Gauss.


The intensity fluctuations can be described in various ways, e.g.  
by the Heisenberg-Langevin approach, a Fokker-Planck analysis, the quantum 
regression theorem, etc.~\cite{book}, each provides their own insights
to the classical and quantum fluctuations in the system under investigation. 
Intensity correlations and sub-Poisson statistics 
were observed with two lasers in EIT regime~\cite{Alzar2,Xiao}. 
In nonlinear magneto-rotation~\cite{budker02rmp}, 
quadrature squeezing theoretically predicted~\cite{squeezing} 
has been experimentally demonstrated~\cite{Lvovsky}.

In the present experiment, 
we have two beams of strong laser radiation with different polarizations
interacting with the atomic system (as shown in Fig.~1a).  
There are two mechanisms by which 
the intensity fluctuations can be generated in an atomic medium. 
One is related to the phase fluctuations of laser field (a linear effect)
\cite{linear-noise} and the second is four-wave mixing of vacuum modes 
(a non-linear effect) \cite{4wm-noise}. 
It was shown that the efficiency of phase noise (PhN) conversion 
to intensity noise (IN) is smaller 
at higher intensity, when non-linear processes are
dominant. The PhN-to-IN conversion phenomenon was used for noise
spectroscopy of resonance medium~\cite{linear-noise}. 
Noise spectroscopy of EIT is studied in~\cite{noise-spetr}. 
Also excess noise was observed
under EIT conditions. We studied correlations on delay time
and dependence of the maximal correlation on two-photon
detuning. In our case optical density has been large 
and noise spectrum is defined by the shape of spectral hole due to
one-photon saturation~\cite{saturation}. 

Physical insight into this process is gained on the basis 
of a simplified approach treating the laser fields clasically,
using the density matrix for treating the resonance atomic responce, 
and taking 
into account propagation that are very essential for the problem. 
The laser beams are in the resonance three-level medium 
as depicted in Fig.~1. 


Indeed, the polarizations and coherence in the three-level system are 
given \cite{book} by
\beq
\rho_{cb} = -{\G_{ca} + \G_{ab}\over2\G_{ca}\G_{ab}}{\W_1\W_2\over
\G_{cb} + \ds{{|\W_2|^2\over\G_{ab}} +
{|\W_1|^2\over\G_{ca}}}
},\\
\rho_{ab} = -i{n_{ba}\W_1 + \rho_{cb}\W_2\over\G_{ab}},
\;\;\;
\rho_{ca} =  i{n_{ca}\W_2 + \rho_{cb}\W_1\over\G_{ca}}
\eeq
where
$\G_{ab} = \g_{ab} + i(\w_{ab}-\nu_1)$;
$\G_{ca} = \g_{ca} - i(\w_{ac}-\nu_2)$;
$\G_{cb} = \g_{cb} + i(\w_{cb}-\nu_1 + \nu_2)$;
where $\w_{\alpha\beta}$ are the atomic frequencies, and 
$\nu_{1}(t) = \nu_2(t)$ are the instant frequency of laser radiation 
in both beams having orthogonal polarizations.  
Practically all population is destributed between levels $b$ and $c$, 
$n_b=n_c \simeq 1/2$, no population is in level $a$, $n_a=0$. 
The equations for field propagation are
\be
\frac{\partial\W_1}{\partial{z}}=-i\eta_b\rho_{ab},
\;\;\;
\frac{\partial\W_2}{\partial{z}}=-i\eta_c\rho_{ac},
\ee
where $\eta_b=\nu_1 N \wp_{b}/(2\epsilon_0 c)$, 
$\eta_c=\nu_2 N \wp_{c}/(2\epsilon_0 c)$ are the coupling constants,
$N$ is the density of medium,
$\epsilon_0$ is the permitivity of the vacuum, and $c$ is the
speed of light in vacuum.  

Depending on the one- and two-photon 
detunings the amplitude fluctuations for fields occurs in phase or $\pi$ out of
phase. When two fields are in the resonance with atomic transitions, 
phase fluctuation of laser field leads to detuning from the resonance, but
one-photon detunings for both fields are the same, so the change of intensities
are in the same direction, and the fields are correlated. When the detuning 
is included, then the change of absorption is different for these two beams. 
One field frequency is tuning closer to the resonance but another is tuning 
out of resonance. The last gives rise for anti-correlations. The time scale 
to establish correlation or anti-correlation between intensities of laser 
beams depends on long-lived coherence relaxation between magnetic sublevels. 
An intersting result is that the spectral width of correlation function 
is more tolerant to the power broadening giving us 4 times 
norrower peak than the EIT width. 
The second mechanism, using four-wave mixing, 
gives us a nonlinear contribution 
to the intensity fluctuations. Let us note that both mechanisms leads to 
correlation at the EIT condition; correlated intensities can be also 
considered from the point of view of matched pulses~\cite{harris}. 

A useful theoretical approach for this situation is 
the technique developed for the CEL~\cite{Scully,schleich}. 
The system is practically the same, 
but instead of V-scheme we have here Lambda-system, 
and the basic processes, described in term of the CEL analysis,
is locking and unlocking. The basic equation describing the obtain results 
for the angle $\theta$, which is the phase deference of two radiation modes,
is given by
\be
\dot\theta = a - b \sin\theta + {\cal F}(\theta) 
\ee
where  ${\cal F}(\theta) = 
\cos{\theta\over 2}F_-(\theta) + \sin{\theta\over2}F_+(\theta)$, 
$F_\pm(\theta)$ are the Langevin forces.  
When two modes are locked their intensities 
start correlating~\cite{schleich}. 

We performed numerical simulations by using the quantum regression 
theory taking the same values of the parameters as in the
current experiments and the results will be published 
elsewhere~\cite{upcoming}. 

The the correlation peak in Fig.~1b is quite narrow, 
just $\tau = 18$ ns. The corresponding frequency
spectrum has a width 18 MHz that is an order more then EIT width
1.2 MHz. We have observed the noise spectrum of laser radiaiton after the cell
directly with a spectrum analyzer. 
Probably correlated spectral components of coupled
optical fields find appropriate degenerated $\Lambda$ systems.
The magnetic field changes correlations between the waveforms. In
Fig.~1c one can see the anti-correlation peak and some fluctuations
near zero level ($B = -0.47$ Gauss). On the slope of the EIT
resonance phase noise should be more efficiently converted to the
intensity noise. In Figure~2 one can see that  the noise on the slope
of the EIT resonance, more pronounced to compare with at the EIT resonance.

In conclusion, we report an experimental observation of intensity
correlations and anti-correlations of coupled fields in a dense
Rb vapor where lower level coherence is created between
Zeeman sub-levels by two laser beams with orthoganal polarizations. 
Intensity fluctuations induced by resonant medium  
are correlated under resonance EIT condition and anti-correlated
at some value of two photon detuning.
Narrow correlation and anti-correlations peaks are associated with
frequencies above EIT width and natural optical width. Dependence
of correlations on magnetic field (two -photon detuning) show
resonance behavior. The resonances are near 4 times narrower 
than the width of the observed EIT resonances. 

Correlation properties of coupled fields in $\Lambda$
scheme can be used to reduce noise and improve performance of EIT
based atomic clocks and magnetometers.  
The PhN-to-IN conversion is an important physical process limiting 
the accuracy. In EIT atomic clock and magnetometers \cite{clock},
it is possible to avoid the contribution of 
the atomic medium induced excess intensity noise. In $\Lambda$-EIT 
we demonstrated that waveforms of transmitted
optical field can be strongly correlated. If one would detect coupled 
fields separately by two independent photo-detectors and then
subtract signals, the noise would be reduced to photonic
shot noise or even less  (sub-Poissonian photon statistics of
coupled optical fields). 




We thank G. Ariunbold, S.E. Harris, A. Patnaik, Z.E. Sariyanni, A.S.Zibrov 
for useful and fruitful 
discussions, V.V.Vasiliev for his help with external cavity laser, 
H.Chen for his help with Lab View and gratefully acknowledge the support 
from the Office of 
Naval Research, the Air Force Research Laboratory (Rome, NY), Defense 
Advanced Research Projects Agency-QuIST, Texas A$\&$M University 
Telecommunication and Information Task Force (TITF) Initiative, and 
the Robert A.\ Welch Foundation (Grant No. A-1261).


\end{document}